\documentclass[12pt]{iopart}

\usepackage{graphics}
\begin{document}

\title[QCD Matter Thermalization at RHIC and LHC]{QCD Matter Thermalization
at RHIC and LHC}

\author{Zhe Xu$^1$, Luan Cheng$^2$, Andrej El$^1$, Kai Gallmeister$^3$,
and Carsten Greiner$^1$}

\address{$^1$Institut f\"ur Theoretische Physik,
Goethe-Universit\"at Frankfurt, D-60438 Frankfurt am Main, Germany}
\address{$^2$Institute of Particle Physics, Huazhong Normal University, 
Wuhan 430079, China}
\address{$^3$Institut f\"ur Theoretische Physik, Universit\"at Giessen,
D-35392, Germany}

\ead{xu@th.physik.uni-frankfurt.de}

\begin{abstract}
Employing the perturbative QCD inspired parton cascade, we investigate 
kinetic and chemical equilibration of the partonic matter created in 
central heavy ion collisions at RHIC and LHC energies. Two types of
initial conditions are chosen. One is generated by the model of wounded 
nucleons using the PYTHIA event generator and Glauber geometry.
Another is considered as a color glass condensate.
We show that kinetic equilibration is almost independent on the chosen
initial conditions, whereas there is a sensitive dependence for chemical 
equilibration. The time scale of thermalization lies between 1 and 1.5 fm/c.
The final parton transverse energy obtained from BAMPS calculations is 
compared with the RHIC data and is estimated for the LHC energy.

\end{abstract}


\section{Introduction}
Comparisons between the calculated elliptic flow $v_2$ from
ideal (viscous) hydrodynamics \cite{H01,LR08} and the measured $v_2$ from 
the experiments \cite{rhicv2} at the BNL Relativistic Heavy Ion Collider (RHIC)
support the creation of a nearly equilibrated quark-gluon system flowing
with a small shear viscosity to entropy density ratio $\eta/s$.
On the other hand, initially produced quarks and gluons are far from
thermal equilibrium due to the subsequent rapid longitudinal expansion.
How quarks and gluons thermalize within a short time scale $\le 1$ fm/c as
assumed in hydrodynamical calculations is an important issue. This is
not only because the thermalization time scale has to be theoretically
determined, but also because the mechanism that drives the system toward 
equilibrium should also respond to the smallness of the $\eta/s$ value
and the buildup of the collective flow of the quark gluon plasma (QGP).

To investigate thermalization and collectivity
in a spatially expanding particle system such like the situation in
ultrarelativistic heavy ion collisions, a 3+1 dimensional parton cascade 
Boltzmann Approach of MultiParton Scatterings (BAMPS) \cite{XG05} is developed.
In this talk we will first review the role of perturbative QCD (pQCD) 
gluon bremsstrahlung ($gg\leftrightarrow ggg$) in thermal equilibrium and
in flow buildup. Second, new results on the initial condition dependence
of thermalization and decrease of the transverse energy are given.
Previous studies can be found in \cite{vienna,EXG08}.

We mention that coherent quantum effects like color 
instabilities \cite{instab} may play a role in isotropization of particle
degrees of freedom at the very initial stage where the matter is super dense.
However, more quantitative studies are needed to determine their significance
on the true thermal equilibration in the expanding quark gluon matter.

\section{Parton Cascade BAMPS}
BAMPS solves the Boltzmann equation for partons with pQCD interactions, 
which include, for the moment, gluon elastic scatterings and gluon
bremsstrahlung and its backreaction. The structure of BAMPS is based on
the stochastic interpretation of the transition rate \cite{XG05},
which ensures full detailed balance for multiple scatterings. The recent 
numerical setup can be found in \cite{XG08v2}. The critical energy density
is set to $e_c=0.6$ ${\rm GeV fm}^{-3}$. Gluons are terminated when the
local energy density is smaller than $e_c$.

The differential cross sections and the effective
matrix elements are given by \cite{biro}
\begin{eqnarray}
\label{cs22}
\frac{d\sigma^{gg\to gg}}{dq_{\perp}^2} &=&
\frac{9\pi\alpha_s^2}{(q_{\perp}^2+m_D^2)^2}\,,\\
\label{m23}
| {\cal M}_{gg \to ggg} |^2 &=&\frac{9 g^4}{2}
\frac{s^2}{({\bf q}_{\perp}^2+m_D^2)^2}\,
 \frac{12 g^2 {\bf q}_{\perp}^2}
{{\bf k}_{\perp}^2 [({\bf k}_{\perp}-{\bf q}_{\perp})^2+m_D^2]}
\Theta(k_{\perp}\Lambda_g-\cosh y)\,,
\end{eqnarray}
where $g^2=4\pi\alpha_s$. ${\bf q}_{\perp}$ and
${\bf k}_{\perp}$ denote the perpendicular component of the momentum
transfer and of the radiated gluon momentum in the center-of-mass
frame of the collision, respectively. $y$ is the momentum rapidity of
the radiated gluon in the center-of-mass frame, and $\Lambda_g$ is the
gluon total mean free path, which is calculated self-consistently \cite{XG05}.
The interactions of the massless gluons are screened by a Debye mass
$m_D^2=\pi d_G \,\alpha_s N_c \int d^3p /(2\pi)^3 \, f / p$
where $d_G=16$ is the gluon degeneracy factor for $N_c=3$.
$m_D$ is calculated locally using the gluon density function $f$
obtained from the BAMPS simulation.

The suppression of the bremsstrahlung due to the 
Landau-Pomeranchuk-Migdal (LPM) effect is taken into account within 
the Bethe-Heitler regime employing the step function in equation (\ref{m23}).
The mean free path $\Lambda_g$ serves here as an infrared regulator, which
leads to a lower cutoff for the transverse momentum of the radiated 
(or absorbed) gluon. Compared to elastic collisions, the collision angles
in a bremsstrahlung become larger due to the additional regulator \cite{XG05}.
This makes the $gg\leftrightarrow ggg$ processes much more efficient for
kinetic equilibration.
Only for large value of $\Lambda_g\sqrt{s}$, in an ultrahigh energy collision
for instance, the cutoff is small and the distribution of collision angles
is forwards directed, which is similar to the elastic case \cite{XG07}.

\section{The role of $gg\leftrightarrow ggg$ in thermalization and in flow
buildup}
The left panel of Fig. 1 shows the momentum
isotropization obtained from BAMPS calculations with a constant QCD coupling
of $\alpha_s=0.3$ for a central Au+Au collision at $\sqrt{s_{NN}}=200$ GeV.
\begin{figure}[t]
\label{trate}
\begin{center}
\resizebox{0.47\textwidth}{!}{
  \includegraphics{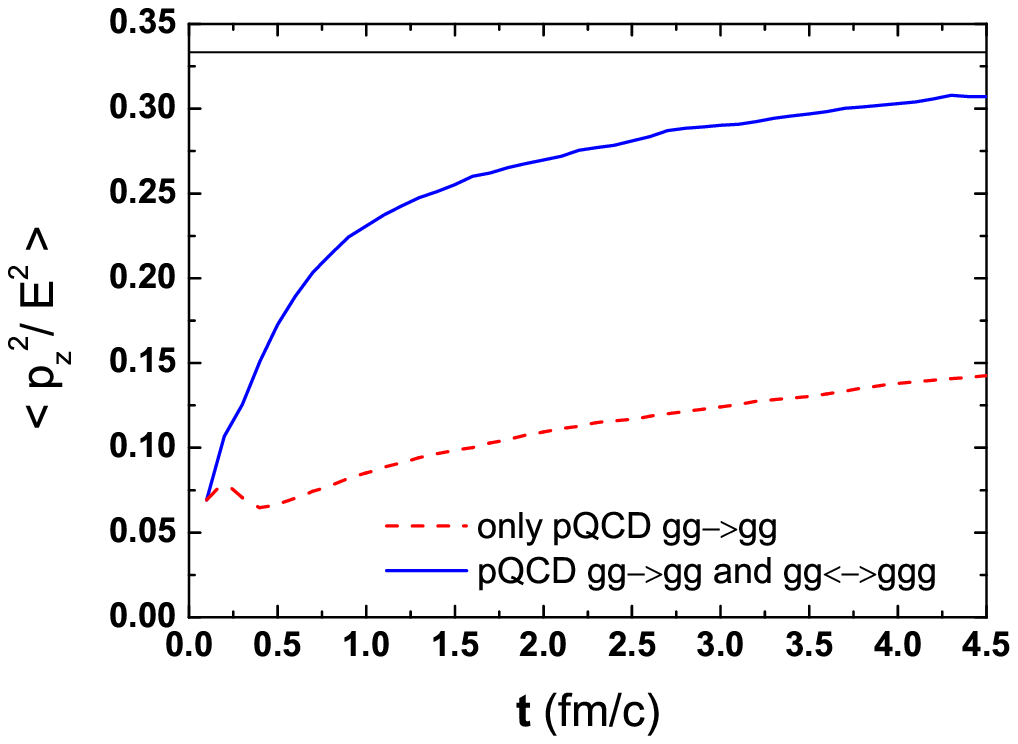}
}
\hspace{\fill}
\resizebox{0.48\textwidth}{!}{
  \includegraphics{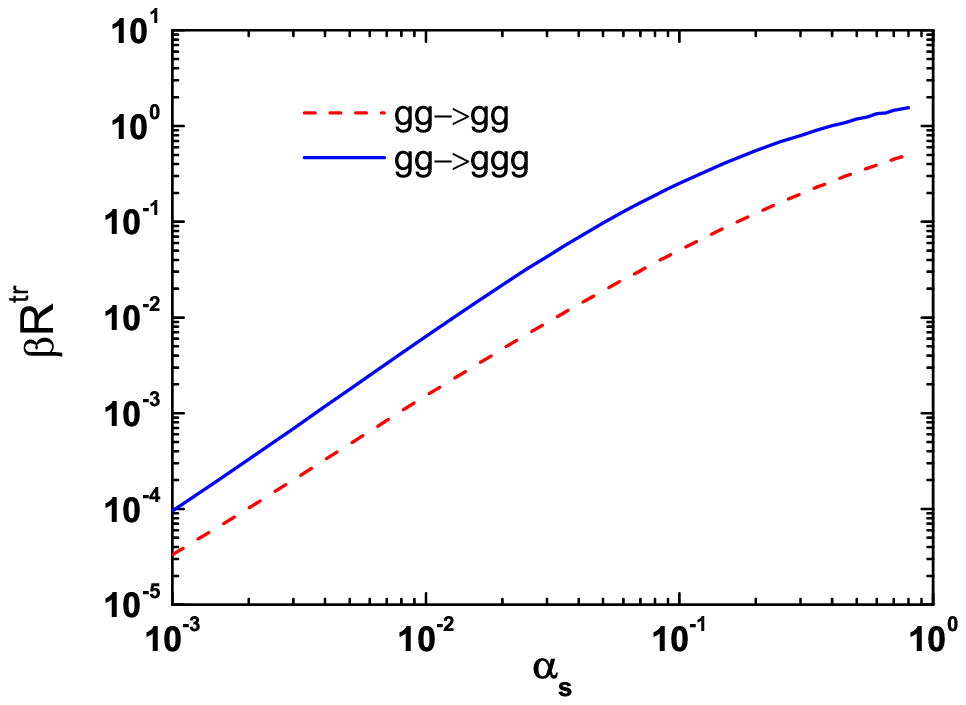}
}
\caption{(color online) Left panel: momentum isotropization in the 
central region. Right panel: transport collision rate scaled by temperature.
}
\end{center}
\end{figure}
The initial gluon distributions are taken as an ensemble of minijets 
with transverse momenta greater than $1.4$ GeV \cite{XG07}, produced via 
semihard nucleon-nucleon collisions with the Glauber geometry.
The results of $\langle p_Z^2/E^2 \rangle$ are extracted at the collision 
center within the space time 
rapidity $-0.2 < \eta_s < 0.2$ and transverse radius $x_T < 1.5$ fm.
Whereas the only elastic pQCD scatterings cannot drive the system
toward equilibrium, the inclusion of the gluon bremsstrahlung and its 
backreaction enormously speeds up the kinetic equilibration.

The large-angle distribution in bremsstrahlung is the reason for its
efficiency in thermalization. Quantitatively we demonstrated 
that the contributions of different processes to momentum 
isotropization are quantified by the transport rates \cite{XG07}
\begin{equation}
\label{trate_d}
R^{\rm tr}_i= \frac{\int \frac{d^3p}{(2\pi)^3} \frac{p_z^2}{E^2} C_i[f] -
\langle \frac{p_z^2}{E^2} \rangle \int \frac{d^3p}{(2\pi)^3} C_i[f]}{n\,
(\frac{1}{3}- \langle \frac{p_z^2}{E^2} \rangle)} \,,
\end{equation}
where $C_i[f]$, functional of the gluon density distribution $f(p,x)$,
is the corresponding collision term describing various interactions,
$i=gg\to gg, gg\to ggg, ggg\to gg$, respectively.
The sum of $R^{\rm tr}_i$ and that of particle drift gives the inverse 
of the time scale of momentum isotropization \cite{XG07}, which also
marks the time scale of overall thermalization.
The right panel of Fig. 1 shows the transport collision
rate, scaled by temperature $T=1/\beta$, for elastic $gg\to gg$ scattering
and bremsstrahlung $gg\to ggg$, respectively.
$R^{\rm tr}_{gg\to ggg}$ is a factor of $3-5$ larger than
$R^{\rm tr}_{gg\to gg}$ over a range in the coupling constant $\alpha_s$
from $10^{-3}$ to $0.8$, which demonstrates the essential role of the
bremsstrahlung in thermal equilibration. 

In addition, the shear viscosity to the entropy density ratio $\eta/s$ 
is inversely proportional to the sum of transport
collision rate \cite{XG08}. Thus $gg\leftrightarrow ggg$ processes
significantly decrease the $\eta/s$ value. We also found \cite{XGS08} that
$gg\leftrightarrow ggg$ processes build up large elliptic flow $v_2$
observed at RHIC. When $\alpha_s=0.3-0.6$ is chosen, the calculated $v_2$ 
from BAMPS is comparable with the experimental data and the extracted 
QGP $\eta/s$ lies between 0.15 for $\alpha_s=0.3$ and 0.08 for $\alpha_s=0.6$.

\section{Initial conditions: model of wounded nucleons and 
color glass condensate}
To study the initial condition dependence of thermalization we choose
two different types of parton initial conditions in nucleus-nucleus collisions.
One is based on the model of wounded nucleons with the Glauber 
geometry \cite{XG05}. A nucleus-nucleus collision is considered as sequent 
binary nucleon-nucleon collisions. To obtain the parton momentum 
distribution in a proton-proton collision we employ the PYTHIA event
generator \cite{pythia} and turn down the function of the parton fragmentation.
On shell quarks and gluons are produced either by (semi)-hard parton-parton
collisions or by the associated initial and final state radiations.
Soft partons produced are not included to cascade calculations. To obtain 
initial conditions 
in a central Au+Au (Pb+Pb) collision the parton production in a
p+p collision is scaled by a number of binary collisions $N_{\rm bin}$, 
which is set to be $N_{\rm bin}=1000$. The positions of the initial partons
are determined according to the nuclear overlapping density within the
Glauber geometry using the Woods-Saxon profile \cite{XG07}. We note that
exact fractions of proton-proton, proton-neutron, and neutron-neutron 
collisions and shadowing effects in a nucleus-nucleus collision will be 
taken into account in a forthcoming paper.

Another type of initial conditions is considered as the color glass
condensate (CGC) within a $k_T$-factorization KLN 
approach \cite{KLN04,HN04,dumi}
\begin{equation}
\label{kln}
\frac{dN_g}{d^2r_T dy}=\frac{4N_c}{N_c^2-1}\int^{p_T^{max}} 
\frac{d^2p_T}{p_T^2} \int d^2k_T \, \alpha_s 
\phi_A(x_1,{\bf k}_T^2;{\bf r}_T)\,
\phi_B(x_2,({\bf p}_T-{\bf k}_T)^2;{\bf r}_T)\,.
\end{equation}
Applying such initial conditions to ideal hydrodynamic calculations
final hadronic yields and their spectra measured at RHIC are well 
reproduced \cite{HN04}. 
In addition, CGC initial conditions give larger initial eccentricity than
the Glauber-type ones \cite{dumi}. This leads to larger elliptic flow in 
noncentral nucleus-nucleus collisions \cite{LR08}. The initial condition
dependence on $v_2$ from BAMPS calculations will be investigated in
the near future. In this work, the unintegrated gluon distribution 
$\phi(x,k_T^2;{\bf r}_T)$ is taken from Ref. \cite{dumi}, which,
compared to the original KLN approach, gives
a smooth transition from the saturation, $\phi(k_T^2)=const.$, towards
the DGLAP regime, $\phi(k_T^2)\sim 1/k_T^2$. We set the prefactor for 
$\phi(x,k_T^2;{\bf r}_T)$ so that the total gluon energy is $80\%$ of 
$\sqrt{s}=200(5500)$A GeV at RHIC(LHC), the same as obtained for quarks 
and gluons together in the wounded nucleons model.

\section{Energy decrease and thermalization at RHIC and LHC: dependence
on initial conditions}
The left panel of Fig. 2 shows the momentum rapidity distribution
of transverse energy of initial quarks and gluons produced in wounded 
nucleons model (wn) and CGC approach in a central Au+Au collision at RHIC.
\begin{figure}[ht]
\label{etR}
\begin{center}
\resizebox{0.48\textwidth}{!}{
  \includegraphics{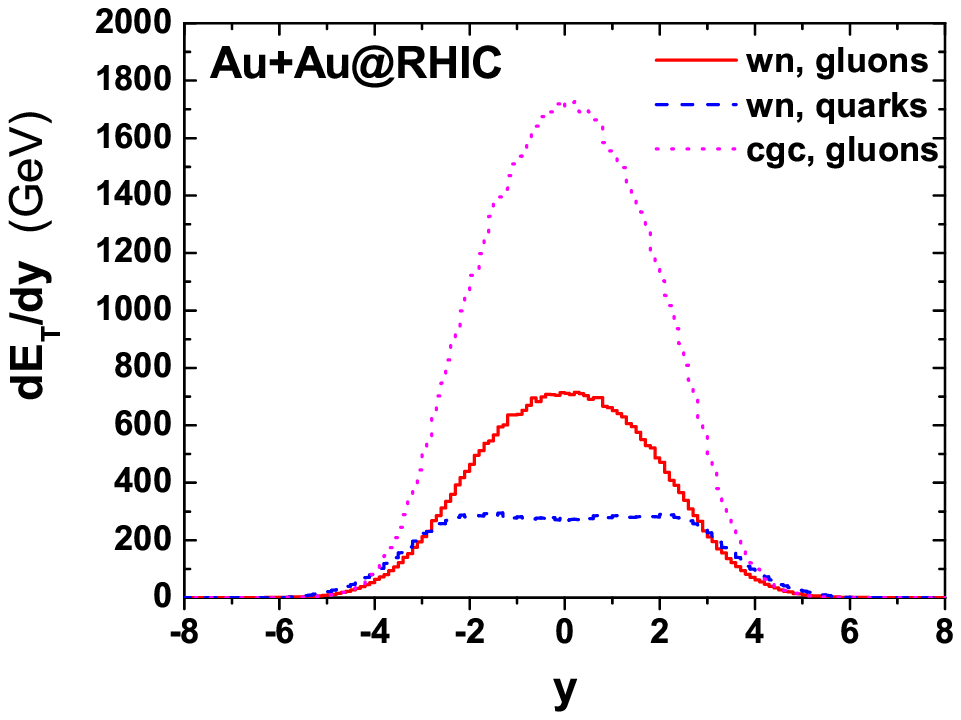}
}
\hspace{\fill}
\resizebox{0.48\textwidth}{!}{
  \includegraphics{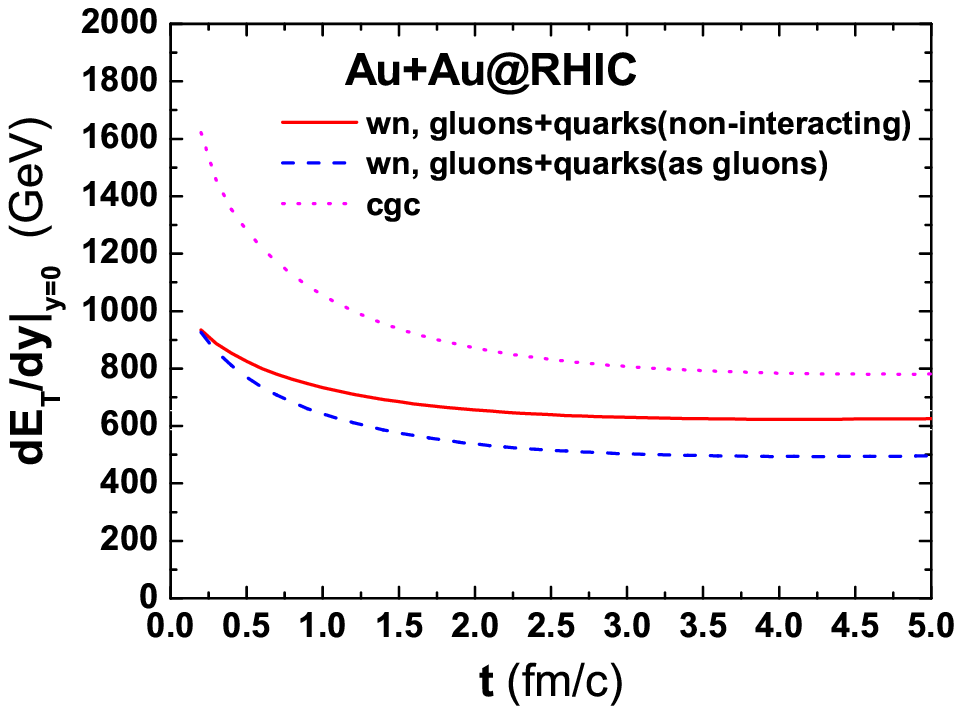}
}
\caption{(color online) Left panel: momentum rapidity distribution of
transverse energy of initial quarks and gluons in a central Au+Au 
collision at RHIC. Right panel: time evolution of the total transverse
energy at midrapidity.
}
\end{center}
\end{figure}
The transverse energy of CGC gluons is larger than that of quarks and gluons
from wn model over a wide range of rapidity except for at large 
rapidity $|y| > 4$. The difference stems from the different approach for
the production of gluons with low transverse momentum: Whereas the CGC
gives a moderate yield (even though suppressed due to the saturation), 
no production is expected in the wn model below a scale of $p_T=1.4-2$ GeV, 
below which the pQCD parton scatterings is no longer valid.

The right panel of Fig. 2 shows the decrease of the transverse 
energy at midrapidity calculated from BAMPS simulations using $\alpha_s=0.3$.
For both types of initial conditions a formation time for each parton is
introduced as $\tau_0=0.15 \cosh y$ fm/c, where 0.15 fm/c is the overlapping
time of a Au+Au collision. The solid curve depicts the result
when quarks in the wn model do not interact, while the dashed curve depicts
the result when quarks interact as strong as gluons. The implementation
of real pQCD quark dynamics in BAMPS calculations is in progress. We expect
that the true final transverse energy of quarks and gluons will lie between
the final values of the solid and dashed curve and thus will be 
comparable with the exprimental data including hadronic and electromagnetic 
components, $dE_T/dy|_{y=0}=620\pm 33$ GeV \cite{star}.

The final transverse energy of CGC gluons is about $30\%$ larger than
the experimental data due to the larger initial value compared with that
in the wn model, although the $E_T$ decrease of CGC gluons is slighly
stronger. To match the experimental data, a smaller critical energy density 
$e_c$ has to be taken to have a later kinetic freezeout, or a larger coupling
$\alpha_s$ has to be assumed, which will lead to a smaller $\eta/s$ of
the QGP. Although the latter will contradict the findings from recent viscous
hydrodynamic calculations \cite{LR08}, new analyses with experimental data 
show that the extracted $\eta/s$ with CGC as the initial conditions is 
indeed smaller than that with the Glauber-type approach \cite{snellings-bai}. 
Further comparisons between different extraction models have to be made.

The local momentum isotropization, $\langle p_Z^2/E^2 \rangle (t)$,
is shown in the left panel of Fig. 3.
\begin{figure}[ht]
\label{equR}
\begin{center}
\resizebox{0.48\textwidth}{!}{
  \includegraphics{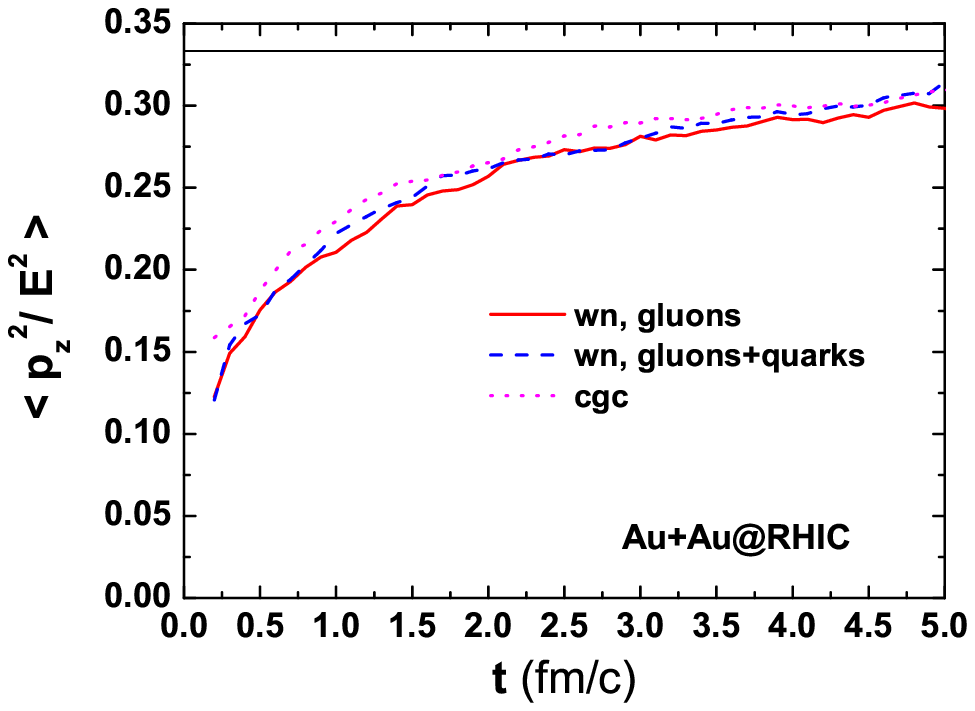}
}
\hspace{\fill}
\resizebox{0.46\textwidth}{!}{
  \includegraphics{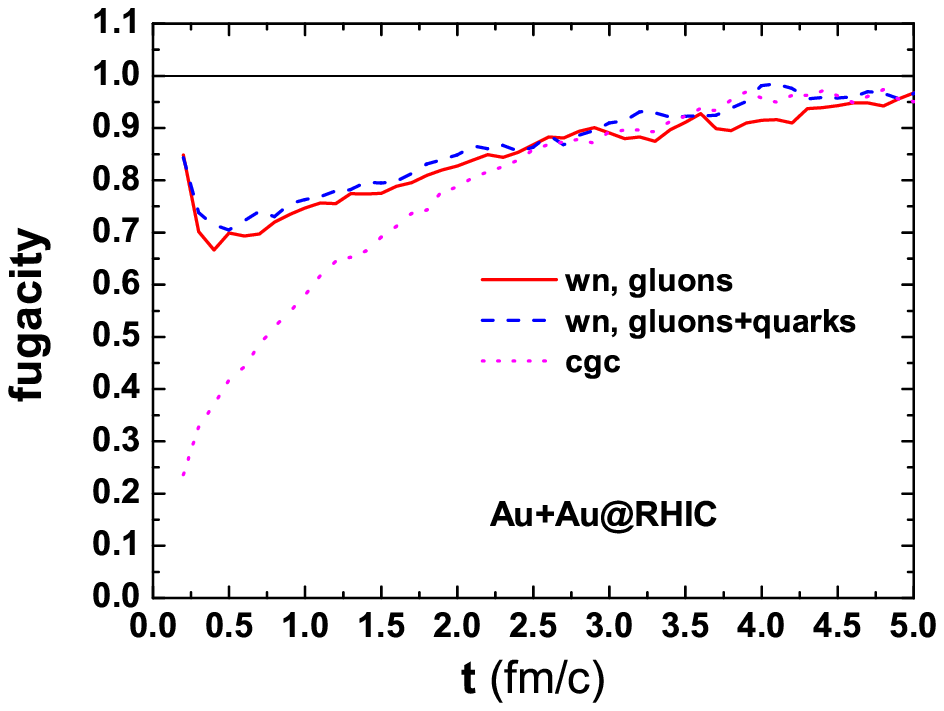}
}
\caption{(color online) Momentum isotropization (left) and time
evolution of the fugacity (right) in the central region in central
Au+Au collisions at RHIC energy.
}
\end{center}
\end{figure}
The solid curve depicts the momentum isotropization of gluons in the wn 
model, while the dashed curve depicts the result including quarks that
are assumed to interact as gluons. Both curves are almost identical. This
indicates that the sum of the transport collision rates $R^{tr}$
(\ref{trate_d}), which determines momentum isotropization \cite{XG07},
is the same in the QCD matter with or without quarks. 

The momentum isotropization of the CGC gluons (dotted curve) is slightly
faster than that of partons in the wn model. Since the initial energy 
density of the CGC gluons is larger (see the left panel of Fig. 2),
the local effective temperature $T=e/(3n)$ is also larger than that in 
the wn model. This leads to a larger $R^{tr}$ , because $R^{tr}$ is 
approximately proportional to $T$ \cite{XG07}.
In any case, the momentum isotropization with the initial conditions from
the wn and CGC approach is not much different from that with minijets
initial conditions shown in the left panel of Fig. 1. The
time scale of the momentum isotropization (kinetic equilibration) is 
about 1.5 fm/c.

The chemical equilibration is described by the time evolution of 
the fugacity $\lambda(t)=n(t)/n_{eq}(t)$, where $n$ is the local particle
density and $n_{eq}=d_G T^3/\pi^2$ is the value at thermal equilibrium.
The chemical equilibration shown in the right panel of Fig. 3
is quite different between the wn and CGC approach. Whereas quarks and gluons
in the wn model is initially almost in chemical equilibrium, the CGC gluons
need a time of 2 fm/c to achieve the chemical equilibrium.

The rapidity distribution of the initial transverse energy in central Pb+Pb
collisions at Large Hadron Collider (LHC) is shown in the left panel of 
Fig. 4.
\begin{figure}[ht]
\label{etL}
\begin{center}
\resizebox{0.48\textwidth}{!}{
  \includegraphics{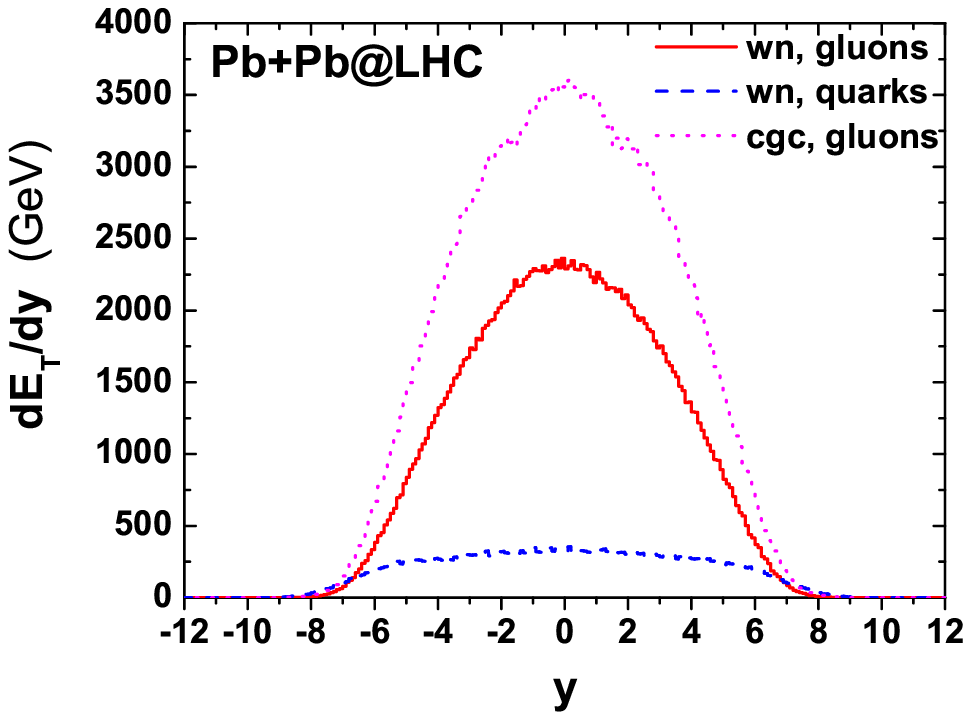}
}
\hspace{\fill}
\resizebox{0.48\textwidth}{!}{
  \includegraphics{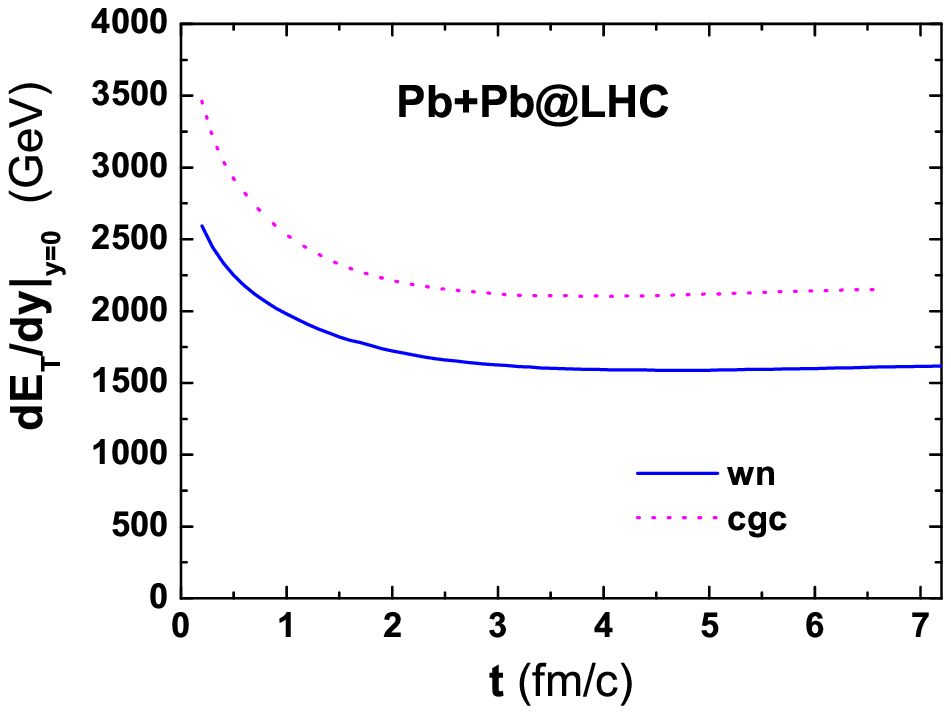}
}
\caption{(color online) Same as Fig. 2 but for central Pb+Pb
collisions at the LHC energy $\sqrt{s}=5500$A GeV with $\alpha_s=0.2$.
}
\end{center}
\end{figure}
The difference between two types of initial conditions is smaller compared
to the case at RHIC, because the saturation scale $Q_s$ of the CGC becomes
larger at higher energy collisions and gluon production below $Q_s$ is
suppressed.

We perform BAMPS calculations using $\alpha_s=0.2$ and the formation time
$\tau_0=0.15 \cosh y$ fm/c.
The final transverse energy at midrapidity is estimated to be between 1620
and 2150 GeV (see the right panel of Fig. 4). The notion
``wn'' in the figure implies calculations including quarks, which are
assumed to behave as gluons. Note that the choices for $\alpha_s$ and 
$\tau_0$ are crucial assumptions. With larger $\alpha_s$ and smaller $\tau_0$
the decrease of the transverse energy becomes stronger.

Figure 5 shows the kinetic (left panel) and chemical (right panel)
equilibration at the collision center of central Pb+Pb collisions at LHC.
\begin{figure}[ht]
\label{equL}
\begin{center}
\resizebox{0.48\textwidth}{!}{
  \includegraphics{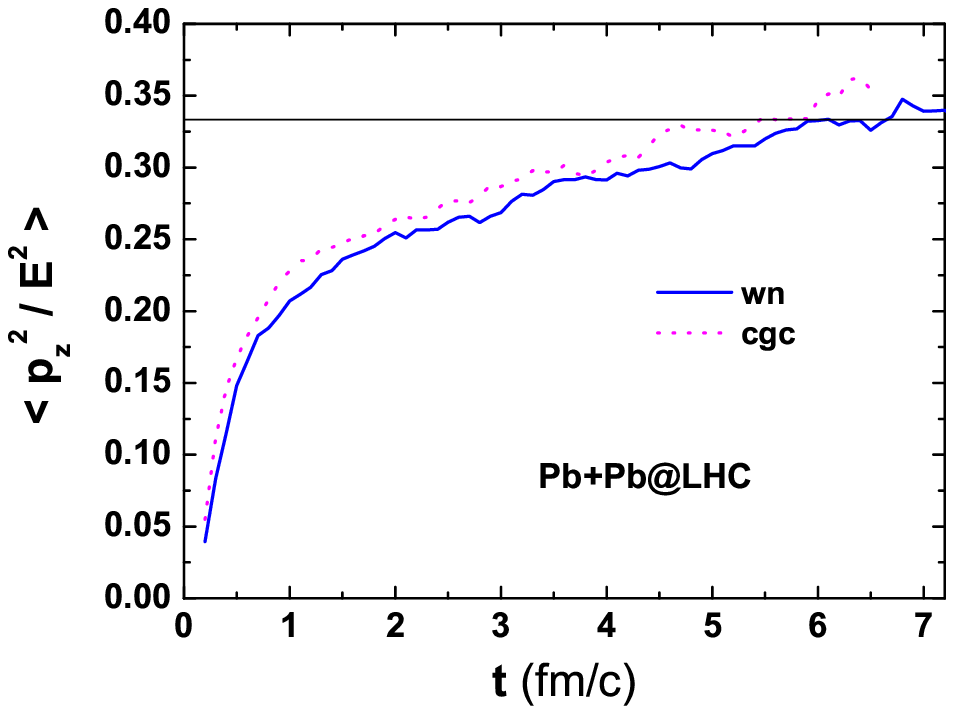}
}
\hspace{\fill}
\resizebox{0.46\textwidth}{!}{
  \includegraphics{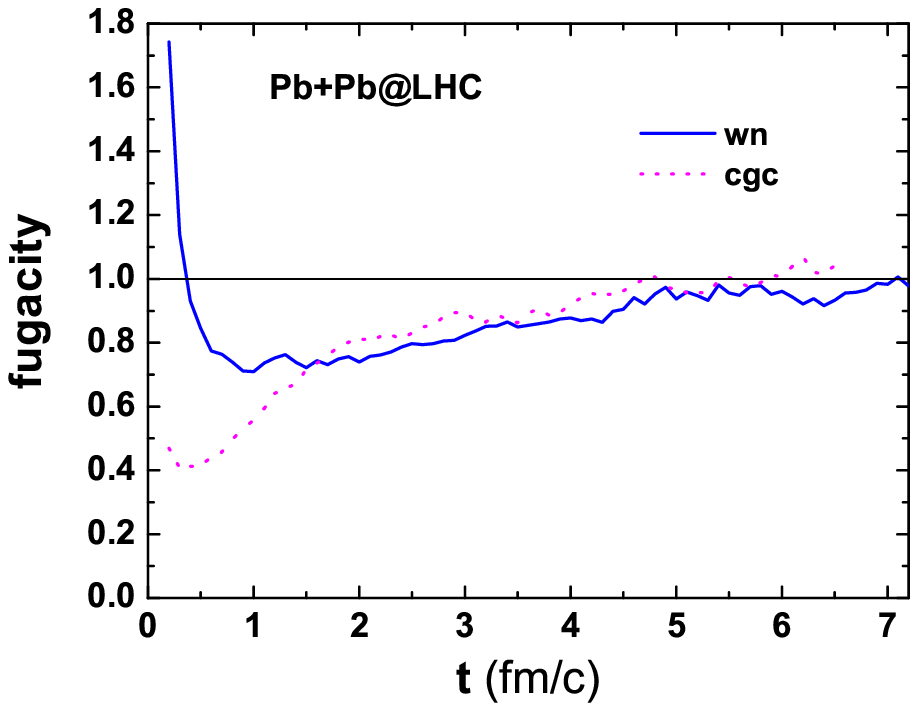}
}
\caption{(color online) Same as Fig. 3 but for central Pb+Pb
collisions at the LHC energy with $\alpha_s=0.2$.
}
\end{center}
\end{figure}
Similar to the case at RHIC, there is no much difference in momentum 
isotropization, whereas the CGC gluons reach the chemical equilibrium
at a later time. Overall thermal equilibrium is expected to be achieved
at about 1 fm/c according to the relaxation formula for 
$\langle p_Z^2/E^2 \rangle$ \cite{XG07}.

\section{Summary}
Employing a parton cascade BAMPS we investigated thermalization of the
QCD matter produced in central nucleus-nucleus collisions at the RHIC and LHC
energy. The transport collision rate is a proper quantity describing kinetic
equilibration and separates contributions from various interactions.
We found the QCD inspired bremsstrahlung is the dominant process in
thermal equilibration and in flow buildup.

We studied the dependence of thermalization on initial conditions
within the wounded nucleons model and the CGC approach.
Using the same value of $\alpha_s$ the momentum
isotropization is almost independent on the initial conditions, whereas
CGC gluons reach chemical equilibrium at a later time than partons in the 
wn model. On the other hand, the final transverse energy of CGC gluons
is higher than that in the wn model due to their initial difference.
A larger $\alpha_s$ for CGC gluons will accelerate the decrease of the
transverse energy to reach a comparable value with the experimental data.
This will also accelerate thermalization of CGC and will lead to a smaller
$\eta/s$ ratio. Detailed study including real quark dynamics and connecting
the initial condition dependence of the buildup of elliptic flow $v_2$ 
will be done in the near future.

\ack 
The authors are grateful to H.-J.~Drescher and A.~Dumitru for fruitful 
discussions and to the Center for the Scientific Computing (CSC) at
Frankfurt for the computing resources. This work was supported by
BMBF, DAAD, DFG, GSI and by the Helmholtz International Center
for FAIR within the framework of the LOEWE program (Landes-Offensive zur
Entwicklung Wissenschaftlich-\"okonomischer Exzellenz) launched
by the State of Hesse.


\section*{References}
\begin{thebibliography}{10}
\bibitem{H01}
Huovinen P {\it et al} 2001 {\it Phys. Lett.} B {\bf 503} 58;
Romatschke P and Romatschke U 2007 {\it Phys. Rev. Lett.} {\bf 99} 172301;
Dusling K and Teaney D 2008 {\it Phys. Rev. } C {\bf 77} 034905;
Song H and Heinz U W 2008 {\it Phys. Rev. } C {\bf 77} 064901.
\bibitem{LR08}
Luzum M and Romatschke P 2008 {\it Phys. Rev. } C {\bf 78} 034915.
\bibitem{rhicv2}
Adler S S {\em et al.} (PHENIX Collaboration) 2003
{\it Phys. Rev. Lett.} {\bf 91} 182301;
Adams J {\em et al.} (STAR Collaboration) 2004
{\it Phys. Rev. Lett.} {\bf 92} 052302;
Adams J {\em et al.} (STAR Collaboration) 2005
{\it Phys. Rev. } C {\bf 72} 014904;
Back B B {\it et al.} (PHOBOS Collaboration) 2005
{\it Phys. Rev. } C {\bf 72} 051901(R);
Adare A {\it et al.} (PHENIX Collaboration) 2007
{\it Phys. Rev. Lett.} {\bf 98} 162301.
\bibitem{XG05}
Xu Z and Greiner C 2005 {\it Phys. Rev. } C {\bf 71} 064901.
\bibitem{vienna}
Xu Z and Greiner C 2006 {\it Eur. Phys. J.} A {\bf 29} 33.
\bibitem{EXG08}
El A, Xu Z, and Greiner C 2008 {\it Nucl. Phys. } A {\bf 806} 287.
\bibitem{instab} 
Mr\'{o}wczy\'{n}ski S 1993 {\it Phys. Lett.} B {\bf 314} 118;
1994 {\it Phys. Rev. } C {\bf 49} 2191;
1997 {\it Phys. Lett.} B {\bf 393} 26;
Arnold P, Lenaghan J, and Moore G D 2003 JHEP {\bf 0308}, 002;
Arnold P {\it et al.} 2005 {\it Phys. Rev. Lett.} {\bf 94} 072302;
Rebhan A, Romatschke P, and Strickland M 2005
{\it Phys. Rev. Lett.} {\bf 94} 102303;
Dumitru A and Nara Y 2005 {\it Phys. Lett.} B {\bf 621} 89;
Schenke B {\it et al.} 2006 {\it Phys. Rev. } D {\bf 73} 125004.
\bibitem{XG08v2}
Xu Z and Greiner C  e-Print: arXiv:0811.2940.
\bibitem{biro} 
Biro T S {\it et al.} 1993 {\it Phys. Rev. } C {\bf 48} 1275.
\bibitem{XG07}
Xu Z and Greiner C 2007 {\it Phys. Rev. } C {\bf 76} 024911.
\bibitem{XG08} 
Xu Z and Greiner C 2008 {\it Phys. Rev. Lett.} {\bf 100} 172301.
\bibitem{XGS08} 
Xu Z, Greiner C, and St\"ocker H 2008 {\it Phys. Rev. Lett.} {\bf 101} 082302.
\bibitem{pythia}
Sjostrand T, Mrenna S, and Skands P 2006 JHEP {\bf 0605} 026.
\bibitem{KLN04}
Kharzeev D and Nardi M 2001 {\it Phys. Lett.} B {\bf 507} 121;
Kharzeev D, Levin E, and Nardi M 2004 {\it Nucl. Phys.} A {\bf 730} 448
[Erratum {\it ibid.} A {\bf 743} 329];
2005 {\it Nucl. Phys.} A {\bf 747} 609.
\bibitem{HN04}
Hirano T and Nara Y 2004 {\it Nucl. Phys.} A {\bf 743} 305.
\bibitem{dumi}
Adil A {\it et al.} 2006 {\it Phys. Rev. } C {\bf 74} 044905.
\bibitem{star}
Adams J {\it et al.} (STAR Collaboration) 2004 {\it Phys. Rev. } C {\bf 70} 054907.
\bibitem{snellings-bai}
Bai Y and Snellings R, these proceedings.
\end{thebibliography}
\end{document}